# Orthogonal Multilevel Spreading Sequence Design

H.M. de Oliveira, R.M. Campello de Souza
CODEC - Communications Research Group
Departamento de Eletrônica e Sistemas - CTG- UFPE
C.P. 7800, 50711-970, Recife-PE, Brazil
E-mail: {hmo,ricardo}@ufpe.br

**ABSTRACT**
Finite field transforms are offered as a new tool of spreading sequence design. This approach exploits orthogonality properties of synchronous non-binary sequences defined over a complex finite field. It is promising for channels supporting a high signal-to-noise ratio. New digital multiplex schemes based on such sequences have also been introduced, which are multilevel Code Division Multiplex. These schemes termed Galois-field Division Multiplex (GDM) are based on transforms for which there exists fast algorithms. They are also convenient from the hardware viewpoint since they can be implemented by a Digital Signal Processor. A new Efficient-bandwidth code-division-multiple-access (CDMA) is introduced, which is based on multilevel spread spectrum sequences over a Galois field. The primary advantage of such schemes regarding classical multiple access digital schemes is their better spectral efficiency. Galois-Fourier transforms contain some redundancy and only cyclotomic coefficients are needed to be transmitted yielding compact spectrum requirements.

1. INTRODUCTION

Digital multiplex usually concerns Time Division Multiplex (TDM). However, it can also be achieved by Coding Division Multiplex (CDM) which has recently been the focus of interest, especially after the IS-95 standardisation of the CDMA system for cellular telephone [QUAL 92]. The CDMA is becoming the most popular multiple access scheme for mobile communication. Classical multiplex increases simultaneously the transmission rate and the bandwidth by the same factor, keeping thus the spectral efficiency unchanged. In order to achieve (slight) better spectral efficiencies, classical CDMA uses waveforms presenting a nonzero but residual correlation. It is well known that spread spectrum sequences provide multiple-access capability and low probability of interception. We introduce here a new and powerful topic on CDMA techniques that can be implemented along with fast transform algorithms.

In this paper a design of spreading spectrum sequences is introduced which is based is based upon Galois field Transforms (GFT) such as the Finite Field Fourier Transform (FFFT) introduced by Pollard [POL 71], so they seem to be attractive from the implementation point of view. The FFFT has been successfully applied to perform discrete convolution and image processing [REE et al. 77, REE&TRU 79], among many other applications. In this paper we are concerned with a new finite field version [CAM et al. 98] of the integral transform introduced by R.V.L. Hartley [HAR 42, BRI 92]. More details can be found in the companion paper "The Complex Finite Field Hartley Transform". Alike classical Galois-Fourier transforms [BLA 79], the Finite Field Hartley Transform (FFHT), which is defined on a Gaussian integer set $GI(p^m)$, contains some redundancy and only the cyclotomic coset leaders of the transform coefficients need to be transmitted. This yields new "*Efficient-Spread-Spectrum Sequences for band-limited channels*". Trade-offs between the alphabet extension and the bandwidth are exploited in the sequel. Tributaries are rather stacked instead of time or frequency interleaved. This paper estimates the *bandwidth compact-ness factor* relatively to Time Division Multiple Access (TDMA) showing that it strongly depends on the alphabet extension. The synchronous spreading provides a null inter-user interference.

Another point to mention is that the superiority of the spreading spectrum sequences is essentially due to the their low implementation complexity.

## 2. NEW SYNCHRONOUS MULTILEVEL SPREADING SPECTRUM SEQUENCES

Given a signal v over a finite field, we deal with the Galois domain considering the spectrum V over an extension field which corresponds to the Finite Field Transform (Galois Transform) of the signal v [BLA 79, CAM *et al.* 98]. Let $v = (v_0, v_1, ..., v_{N-1})$ be a vector of length N with components over GF(q), $q = p^r$. (hereafter the symbol := denotes *equal by definition*). The FFFT of v is the vector $F = (F_0, F_1, ..., F_{N-1})$ of components $F_k \in GF(q^m)$, given by

$$F_k := \sum_{i=0}^{N-1} v_i \alpha^{ki},$$

where $\alpha$ is a specified element of multiplicative order N in $GF(q^m)$. The FFHT of v is $V = (V_0, V_1, ..., V_{N-1})$ of components $V_k \in GI(q^m)$, given by

$$V_k := \sum_{i=0}^{N-1} v_i \, cas_k(\angle \alpha^i)$$

where $\alpha$ is a specified element of multiplicative order N in $GF(q^m)$. The cas(.) is the Hartley "cosine and sine" kernel defined over a finite field [CAM *et al.* 98].

Each symbol in the ground field GF(p) has duration T seconds. Spreading waveforms can be used to implement an N-user multiplex on the extension field $GF(p^m)$ where $N \mid p^m-1$. For the sake of simplicity, we begin with m=1 and consider a (p-1)-channel system as follows. Typically, we can consider GF(3) corresponding to Alternate Mark Inversion (AMI) signalling.

<u>Definition 1</u>. A Galois modulator (Figure 1) carries a pairwise multiplication between a signal $(v_0, v_1, ..., v_{N-1})$, $v_i \in GF(p)$ and a carrier $(c_0, c_1, ..., c_{N-1})$, with $c_i \in GI(p)$. ∎

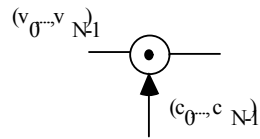

**Fig. 1**. A pictorial representation of a Galois modulator.

A (p-1)-CDM considers, as digital carrier sequences per channel, versions of the cas function $\{cas_i k\}_{k=0}^{p-1}$ over the Galois (complex) field GI(p). The cas (cos and sin) function is defined in terms of finite field trigonometric functions [CAM et al. 98] according to $cas\, k := \cos\, k + \sin\, k$.

**carrier 0:**
$\{cas_0 0 \quad cas_0 1 \quad cas_0 2 \quad ... \quad cas_0(N-1)\}$
**carrier 1:**
$\{cas_1 0 \quad cas_1 1 \quad cas_1 2 \quad ... \quad cas_1(N-1)\}$
...
**carrier j:**
$\{cas_j 0 \quad cas_j 1 \quad cas_j 2 \quad ... \quad cas_j(N-1)\}$
...
**carrier N-1:**
$\{cas_{N-1} 0 \quad cas_{N-1} 1 \quad cas_{N-1} 2 \, ... \, cas_{N-1}(N-1)\}.$

The cyclic digital carrier has the same duration T of an input modulation symbol, so that it carries N slots per data symbol. The interval of each cas-symbol is T/N and therefore the bandwidth expansion factor when multiplexing N channels may be roughly N.

A spread spectrum multiplex is showed in Figure 2: The output corresponds exactly to the finite field Hartley Transform of the "user"-vector ($v_0, v_1, ..., v_{N-1}$). Therefore, it contains all the information about all channels. Each coefficient $V_k$ of the spectrum has duration T/N.

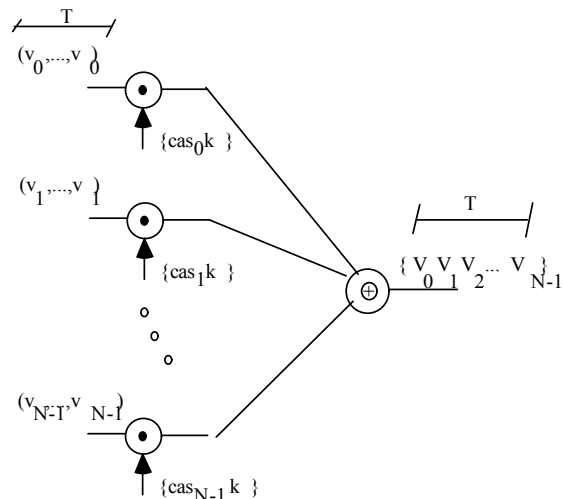

**Fig. 2**. Galois-Field MUX: Spreading sequences.

These carriers can be viewed as spreading waveforms [MAS 95]. An N-user mux has N spreading sequences, one per channel. The requirements to achieve Welch's lower bound according to Massey and Mittelholzer [MAS&MIT 91] are achieved by $\{cas_i k\}_{k=0}^{p-1}$ sequences. The matrix [$\{cas_i k\}$] presents both orthogonal rows and orthogonal columns having the same "energy".

## 3. IMPLEMENTATION OF A SPREADING SPECTRUM SCHEME BY FINITE FIELD TRANSFORMS

### 3.1 Transform Spreading Spectrum

As an alternative and attractive implementation of the spread spectrum system described in the previous section, the spreading can be carried out by a Galois Field Transform (FFFT/FFHT) of length N | $p^m$-1 so the de-spreading corresponds exactly to an *Inverse Finite-Field Transform* of length N.

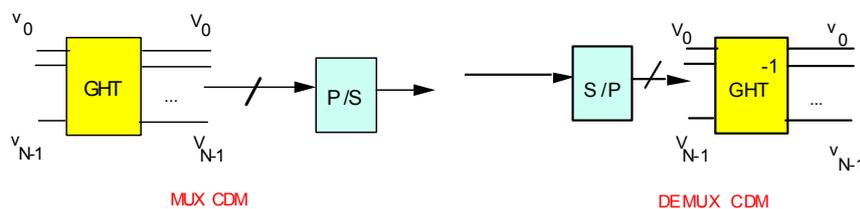

**Fig. 3**. Implementation of Galois Field Transform (GFT) spreading spectrum.

Discrete Fourier transforms (DFT) have been applied in the multicarrier modulation schemes [WEIN&EBER 71] referred to as Orthogonal Frequency Multiplexing (OFDM). Although OFDM presents a very similar block diagram regarding GDM [compare fig. 1 in URI&CAR 99 and Figure 3 above], the true nature of these schemes is quite different. The first one is an analogue, discrete- time, frequency division while the later performs a digital Galois division, i.e.,

|            | OFDM                              | GDM                                    |
|------------|-----------------------------------|----------------------------------------|
|            | *Fourier spectrum*                | *Finite Field Spectrum*                |
| DIVISION   | (frequency domain)                | (Galois domain)                        |

However, they are both orthogonal transform-based schemes that support fast algorithms (Fast Fourier Transforms) being thus very attractive from the complexity viewpoint.

### 3.2 Interpreting Galois-Hartley Transform over GF(5) as Spreading Waveforms

A naive example is presented in order to illustrate such an approach. A 4-channel CDMA over GF(5) can be easily implemented. It is straightforward to see that such signals are neither FDMAed nor TDMAed. The GI(5)-valued cas(.) function is shown on Table I assuming $\alpha$ equals to 2, an element of GF(5) of order four.

Table I. Cas function on GI(5) with $\alpha=2$, an element of order 4.

| $cas_0 0 = 1+j0$ | $cas_0 1 = 1$  | $cas_0 2 = 1$ | $cas_0 3 = 1$  |
|------------------|----------------|---------------|----------------|
| $cas_1 0 = 1+j0$ | $cas_1 1 = j3$ | $cas_1 2 = 4$ | $cas_1 3 = 2j$ |
| $cas_2 0 = 1+j0$ | $cas_2 1 = 4$  | $cas_2 2 = 1$ | $cas_2 3 = 4$  |
| $cas_3 0 = 1+j0$ | $cas_3 1 = 2j$ | $cas_3 2 = 4$ | $cas_3 3 = j3$ |

A 4-channel complex spreading waveforms over GF(5) can be chosen as (Figure 4):

$\{cas_0 k\} = \{1, 1, 1, 1\}$   $\{cas_1 k\} = \{1, j3, 4, j2\}$
$\{cas_2 k\} = \{1, 4, 1, 4\}$   $\{cas_3 k\} = \{1, j2, 4, j3\}$.

If channels number 1, 2, 3, and 4 are transmitting $\{4, 0, 1, 2\}$ respectively, the mux output will be $(2, 3+4j, 3, 3+j)$, which corresponds to

$$\{4,0,1,2\} \otimes \{1,1,1,1\} \equiv 2 \quad \text{mod } 5$$
$$\{4,0,1,2\} \otimes \{1,j3,4,j2\} \equiv 3+4j \quad \text{mod } 5$$
$$\{4,0,1,2\} \otimes \{1,4,1,4\} \equiv 3 \quad \text{mod } 5$$
$$\{4,0,1,2\} \otimes \{1,j2,4,j3\} \equiv 3+j \quad \text{mod } 5$$

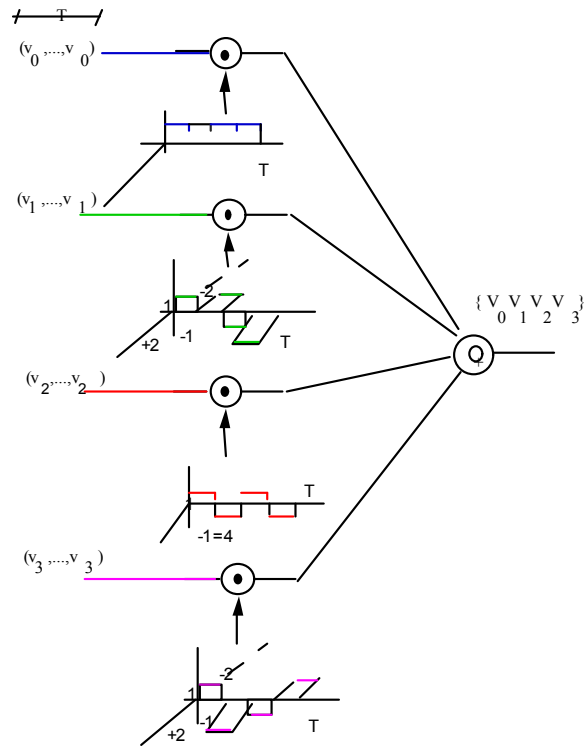

**Fig. 4** Interpreting Galois-Hartley Transform over GF(5) as spreading waveforms.

The digital carriers are defined on a complex Galois field GI(p) where the elements may or not may belong to GF(p), although the original definition [CAM et al. 98] considers -1 as a quadratic non-residue in GF(p). Two distinct cases are to be considered: p=4k+1 and p=4k+3, k integer. For instance, considering j ∈ GF(5) then $2^2 \equiv -1$ (mod 5) so $j = \sqrt{-1} \equiv 2$ (mod 5). Two-dimensional digital carriers then degenerated to one-dimensional carriers. Considering the above example, carriers are reduced to Walsh carriers!

$$\{cas_0 k\} = \{1, \quad 1, 1, 1\} = \{1,1, 1, 1\}$$
$$\{cas_1 k\} = \{1, \quad 1, 4, 4\} = \{1,1,-1,-1\}$$
$$\{cas_2 k\} = \{1, \quad 4, 1, 4\} = \{1,-1,1, -1\}$$
$$\{cas_3 k\} = \{1, \quad 4, 4, 1\} = \{1,-1,-1, 1\}.$$

In the absence of noise, there is no cross talk from any user to any other one, which corresponds to orthogonal carrier case. Considering the above example, carriers are reduced to Hadamard carriers. Therefore, this spreading sequences can be interpreted as some kind of generalisation (multilevel and 2-dimensional) of classical synchronous orthogonal Hadamard spreading spectrum sequences. There is *no gain* when the transform is taken without alphabet extension. However, we have a nice interpretation of CDM based on finite field transforms.

## 4. EFFICIENT-BANDWIDTH CODE-DIVISION MULTIPLE ACCESS FOR BAND-LIMITED CHANNELS

The main title of this section is, apart from the term CDMA, literally identical to a Forney, Gallager and co-workers paper issued more than one decade ago [FOR *et al.* 84], which analysed the benefits of coded-modulation techniques. The large success achieved by Ungerboeck's coded-modulation came from the way redundancy was introduced in the encoder [UNG 82]. In classical channel coding, redundant signals are *appended* to information symbols in a way somewhat analogous to time division multiplex TDM (envelope interleaving). It was believed that introducing error-control ability would increase bandwidth. An efficient way of introducing such ability without neither sacrificing rate nor requiring more bandwidth consists in adding redundancy by an alphabet expansion. This technique is particularly suitable for channels in the narrow-band region. A similar reasoning occurs in the multiplex framework: many people nowadays believe that multiplex must increase bandwidth requirements.

In the present work, the coded-modulation idea [UNG 82] is adapted to multiplex: Information streaming from users is not combined by interleaving (like TDM) but rather by a *signal alphabet expansion*. The mux of users' sources over a Galois Field GF(p) deals with an expanded signal set having symbols from an extension field $GF(p^m)$, m>1. As a consequence, the multiplex of N band limited channels of identical maximal frequency B leads to *bandwidth requirements less than* N B, in contrast with TDMAed or FDMAed signals. It is even possible multiplexing without requiring bandwidth expansion or multiplexing with bandwidth compression provides that the signal-to-noise ration is high enough.

So far we have essentially considered Finite Field Transforms from GF(p) to GF(p). Extension fields can be used and results are much more interesting: The Galois-Field Division Multiple Access schemes. The advantage of the new scheme named GDMA over FDMA / TDMA regards its higher spectral efficiency.

The new multiplex is carried out over the Galois domain instead of the "frequency" or "time" domain. As an attractive implementation, the multiplex can be carried out by a Galois Field Transform (FFFT/FFHT) so the DEMUX corresponds exactly to an Inverse Finite-Field Transform of length N | $p^m$-1. Transform-based multiplexes by spreading spectrum perform as follows. First, the Galois spectrum of N-user GF(p)-signals is evaluated. The spectral compression is achieved by eliminating the redundancy: only the leaders of cyclotomic cosets are transmitted.

Figure 5 exhibits a block diagram of transform-based multiplexes. First, the Galois spectrum of N-user GF(p) signals is evaluated. Demultiplex is carried out (after signal regeneration) first recovering the complete spectrum by "filling" missing components from the received coset leaders. Then, the inverse finite field transform is computed so as to obtain the de-mux signals. Another additional feature is that GDM implementations can be made more efficient if fast algorithms for computing the involved transforms are used.

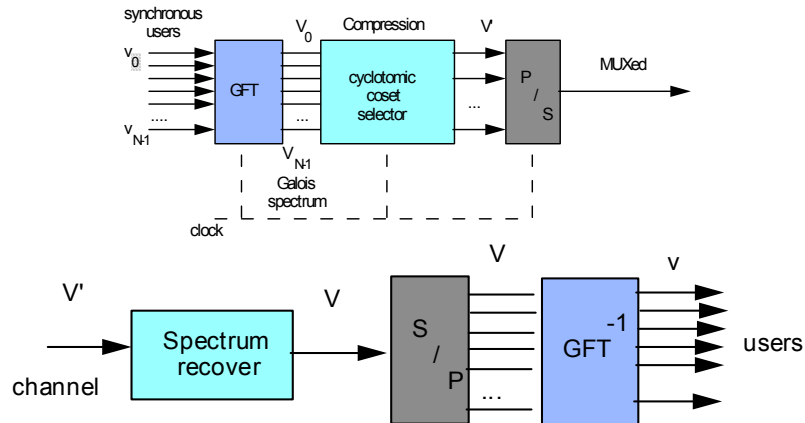

**Fig. 5**. Multiple Access based on Finite Field Transforms.

Suppose that users data are p-ary symbols transmitted at a speed B:=1/T bauds. Let us think about the problem of multiplexing N users. Traditionally the bandwidth requirements will increase proportionally with the number N of channels, i.e., $B_N$=NB Hz.

Thereafter the number of cyclotomic cosets associated with a Galois-Fourier (or Galois-Hartley) spectrum is denoted $\nu_F$ (respectively $\nu_H$). The clock driving multilevel (finite field) symbols is $N/\nu$ times faster than the input baud rate. The bandwidth requirements will be $(N/\nu)B$ instead of $NB$.

Definition 2. The bandwidth compactness parameter $\gamma_{cc}$ is defined as $\gamma_{cc} := N/\nu$. ∎

It plays a role somewhat similar to the coding asymptotic gain $\gamma_c$ on coded modulation [UNG 82].

The processing gain, $G$, of these multilevel CDMA systems is given by the number of cyclotomic cosets, that is, $G = N/\gamma_{cc} = \nu$.

Another point that should be stressed is that instead of compressing spectra (eliminating redundancy), it is possible to use all the coefficients to introduce some error-correction ability. The valid spectrum sequences generate a multilevel block code.

Transform-multiplex, i.e., digital multiplex based on finite field transforms are very attractive compared with FDM/ TDM due to their better spectral efficiency as it can be seen in Table II.

Proposition 1. For an N-user GDMA system over $GI(p^m)$ with $N | p^m - 1$, only a number $\nu = \gamma_{cc}^{-1} N$ (see below) of finite-field transform coefficients are required to be transmitted.

Proof. According to Mœbius' inversion formula, $I_k(q) = \frac{1}{k} \sum_{d | k} \mu(d) q^{k/d}$ gives the number of distinct irreducible polynomials of degree $k$ over $GF(q)$, where $\mu$ is the Mœbius function [McE 87]. Therefore, the number $\nu_F$ of cyclotomic sets on the Galois-Fourier transform $(V_0, V_1, ..., V_{N-1})$ is given by $\nu_c = \sum_{k | N} I_k(p) - 1$.

Since each pair of cosets containing reciprocal roots is clustered, then

$$\nu_e = \frac{\nu_c - (N \bmod 2)}{2} + 1.$$
∎

Table II. A Spectral Efficiency Comparison for Multiple Access Systems.

| | one-user | N-users TDMAed | N-users GDMAed |
|---|---|---|---|
| Transmission Rate | $R_{one-user} = \frac{\log_2 p}{T}$ | $R = \sum_k R_{k-user} = N\frac{\log_2 p}{T}$ bps | $R = \sum_k R_{k-user} = N\frac{\log_2 p}{T}$ bps |
| Bandwidth requirements | $B_N = \frac{1}{T}$ Hz | $B_k = \frac{1}{T/N} = NB_N$ Hz | $B_{GDM} = \frac{1}{T/(\gamma_{cc}^{-1})} = \frac{1}{\gamma_{cc}}(NB_N)$ Hz |
| Spectral Efficiency | $\eta_{one-user} = \log_2 p$ bits/s/Hz | $\eta_{tdm} = \log_2 p$ bits/s/Hz | $\eta_{gdm} = \gamma_{cc} \log_2 p$ bits/s/Hz |

As a rule-of-thumb, the number of cosets in the case of block-length $N=p^m-1$ is roughly given by $v_F \approx \lceil \frac{N}{m} \rceil$ and $v_H \approx \lceil \frac{1}{2} \lceil \frac{N}{m} \rceil + 1 \rceil$, where $\lceil x \rceil$ is the ceiling function (the smallest integer greater than or equal to x). A simple example over $GF(3) \to GI(3^3)$ is presented below: Factoring $x^{26}-1$ one obtains $v_F=10$ and $v_H=6$. For the FFHT, $V_\eta^P = V_{os-\eta P}$ (indexes modulo 26) according to [CAM et al. 98, Lemma 1].

FFFT cosets                FFHT cosets
C0=(0)                     C0=(0)
C1=(1,3,9)                 C1=(1,23,9,25,3,17)
C2=(2,6,18)                C2=(2,6,18,8,24,20)
C4=(4,12,10)               C4=(4,14,10,22,12,16)
C5=(5,15,19)               C5=(5,11,19,21,15,7)
C7=(7,21,11)               C13=(13).
C8=(8,24,20)
C13=(13)
C14=(14,16,22)
C17=(17,25,23).

By way of interpretation, Hartley transforms can be seen as some kind of Digital Single Side Band since the number of cyclotomic cosets of the FFHT is roughly half that of the FFFT. We can therefore say, "*GDM/FFFT is to FDM/AM as GDM/FFHT is to FDM/SSB.*"

Proposition 2. (GDM Gain) The gain on the number of channels GDMed regarding to TDM/FDM over the same bandwidth is N-v, which corresponds to $g_\% = 100(1 - \gamma_{AA}^{-N})\%$.

Proof. The bandwidth gain is $g_{band}=B_{TDM}/B_{GDM}=\gamma_{cc}$ and the saved Bandwidth is given by $B_{TDM} - B_{GDM}$. Calculating how many additional $B_1$-channels (users) can be introduced:

$$(B_{TDM} - B_{GDM})/B_1 = (1 - \frac{1}{\gamma_{AA}})N. \qquad \blacksquare$$

In the previous example, a 26-user GDM furnishes a 20-channels gain ($\approx 77\%$) regarding to TDM.

What can one tell about the alphabet extension? A simple upper bound on the bandwidth compactness factor can be easily derived. The greatest extension that can be used depends on the signal-to-noise ratio, since the total rate cannot exceed Shannon Capacity over the Gaussian channel. Therefore,

$$B_{daj} \, \gamma_{AA} \log_o p \leq B_{daj} \log_o\left(1 + \frac{S}{N}\right) \text{ bps, or } \gamma_{AA} \leq \log_e\left(1 + \frac{S}{N}\right).$$

## 4. CONCLUSIONS

Finite field transforms are offered as a new tool of spreading sequence design. New digital multiplex schemes based on such transforms have been introduced which are *multilevel* Code Division Multiplex. They are attractive due to their better spectral efficiency regarding to classical TDM/CDM which require a bandwidth expansion roughly proportional to the number of channels to be multiplexed. This new approach is promising for communication channels supporting a high signal-to-noise ratio. Although optical fibres are not yet band-limited channels, the new CDM introduced can be adopted on satellite channels or even on cellular mobile communications. Moreover, a Digital Signal Processor (DSP) can easily carry out the Galois-Field Division implementation. A number of practical matters such as imperfect synchronisation, error performance, or unequal user power are left to be investigated. Combined multiplex and error-correcting ability should be investigated. Another nice payoff of GDMA is that when Hartley Finite Field transforms are used, the spreading and de-spreading hardware is exactly the same.

ACKNOWLEDGMENTS
*This first author expresses his deep indebtedness to Professor Gérard Battail whose philosophy had a decisive influence on his way of looking to coding and multiplex. The authors also thanks Professor Paddy Farrell for his constructive criticism.*